\documentclass[a4paper,10pt,journal,twoside,compsoc]{IEEEtran}
\usepackage[numbers,sort&compress,square]{natbib}

\ifCLASSINFOpdf
   \usepackage[pdftex]{graphicx}

\else
 \fi

\usepackage{array}

\usepackage{amsmath,amssymb, amsbsy,bm}
\usepackage{commath}
\usepackage{xcolor}
\usepackage{booktabs}
\usepackage[caption=false,labelformat=simple]{subfig}

\usepackage{url}
\usepackage[normalem]{ulem}
\usepackage{pinlabel}
\usepackage[breaklinks, hidelinks,colorlinks=false]{hyperref}
\usepackage{enumerate}
\usepackage{algorithm}
\usepackage{algpseudocode}

\newcolumntype{L}[1]{>{\arraybackslash}p{#1}}
\newcolumntype{X}[1]{>{\raggedright\let\newline\\\arraybackslash\hspace{0pt}}m{#1}}

\usepackage{xtab,booktabs}
\newcommand{\Son}{SoNSTAR}
\usepackage{xifthen}

\newcommand{\TNR}{\mathrm{TNR}}
\newcommand{\FPR}{\mathrm{FPR}}
\newcommand{\FNR}{\mathrm{FNR}}
\newcommand{\TP}{\mathrm{TP}}
\newcommand{\FP}{\mathrm{FP}}
\newcommand{\TN}{\mathrm{TN}}
\newcommand{\FN}{\mathrm{FN}}
\newcolumntype{L}[1]{>{\raggedright\arraybackslash}p{#1}}

\begin{document}
\title{Sonification of Network Traffic Flow for Monitoring and Situational Awareness} 
\author{Mohamed Debashi and Paul Vickers%
\thanks{M. Debashi and P. Vickers are with the Department of Computer and Information Sciences, Northumbria University, Newcastle upon Tyne, UK, e-mail: \{mohamed.debashi,paul.vickers\}@northumbria.ac.uk.}}
\markboth{Pre-print}%
{Debashi \& Vickers: Sonification of Network Traffic Flow}

\IEEEtitleabstractindextext{%
\begin{abstract}
Maintaining situational awareness of what is happening within a computer network is challenging, not least because the behaviour  happens within computers and communications networks, but also because data traffic speeds and volumes are beyond human ability to process. Visualisation techniques are widely used to present information about the dynamics of network traffic dynamics. Although they provide operators with an overall view and specific information about particular traffic or attacks on the network, they often still fail to represent the events in an understandable way. Also, visualisations require visual attention and so are not well suited to continuous monitoring scenarios in which network administrators must carry out other tasks. Situational awareness is critical and essential for decision-making in the domain of computer network monitoring where it is vital to be able to identify and recognize network environment behaviours.Here we present \Son\ (Sonification of Networks for SiTuational AwaReness), a real-time sonification system to be used in the monitoring of computer networks to support the situational awareness of network administrators. \Son\ provides an auditory representation of all the TCP/IP protocol traffic within a network based on the different traffic flows between between network hosts. \Son\ narrows the gap between network administrators and the cyber environment so they can more quickly recognise and learn about the way the traffic flows within their network behave and change. \Son\ raises situational awareness levels for computer network defence by allowing operators to achieve better understanding and performance while imposing less workload compared to visual techniques. \Son\  identifyies the features of network traffic flows by inspecting the status flags of TCP/IP packet headers. Different combinations of these features define particular traffic events and these these events are  mapped to recorded sounds to generate a soundscape that represents the real-time status of the network traffic environment. Listening to the sequence, timing, and loudness of the different sounds within the soundscape allows the network administrator to monitor the network and recognise anomalous behaviour quickly and without having to continuously look at a computer screen. \\
\end{abstract}
}
\maketitle


\section{Introduction}

Visualisation has been used as a tool for monitoring networks in order to raise situational awareness levels. The static and dynamic visualisation of total and subtotal traffic information (such as bandwidth, speed and current performance) do not allow administrators to acquire a deep and clear understanding of their current network state.  
This is because  attacks  can appear like normal traffic and there are no specific rules that could enable administrators to set their network up to prevent or monitor all attacks. 
Furthermore, each network is unique and what is normal behaviour in one network may be anomalous in another. Therefore, network administrators need tools to provide information in a way which helps them to build a solid understanding of their network environment's behaviour. Unfortunately, existing popular tools such as intrusion detection systems (IDS) and firewalls do not specify why and how certain events happened.  

Visualisation and IDS systems do not provide the protocol flow granularity required to understand how flows are behaving inside a network or why a security system generates false positive alerts or why specific alarms were raised. IDSs detect intrusions and record them to log files which network administrators then have to inspect to try to understand the situation. Many IDSs send an email to the administrator for each intrusion record or incident and the volume of emails increases with the scale of the network. It is quite difficult to understand the relevance of aggregate records when receiving only the alarms for individual intrusion records.  Modern attacks are sophisticated and can involve a range of various. Thus, real time situational awareness is required for an overall understanding of the situation especially when real time intelligence and intuitive solutions are required. The graphical user interfaces (GUIs) of today's nowadays visualisation network monitoring and IDS systems present information very superficially. For example, the time sequence of numbers of intrusions or incidents of the whole traffic domain may be visualised as polygonal charts. The operator may be required to perform many operations to explore detailed information, but in many cases network administrators are too busy to monitor the GUI.  %
Moreover, when using visualisation tools administrators must look at a screen. Loss of concentration, visual fatigue, temporal demand and frustration increase when monitoring a screen for a long period. Extra screens will be required for additional staff. In addition,  the huge volumes of data which need to be processed and presented cannot be visualised in real-time unless data reduction techniques are used.  

Network measurement tools include hardware and software approaches to collect data and analyse traffic at different protocol levels. Network traffic analysers collect real-time data and perform online analysis and the majority of these systems use graphical displays to represent live traffic data.   

Hildebrandt~\cite{Hildebrandt:2014} proposed enhancing visualisation monitoring with  sonification techniques because humans are sensitive to even small changes in the rhythms and sequences of sounds. Sonification may be defined as: 
 \begin{quote}\ldots the use of non-speech audio to convey information. More specifically, sonification is the transformation of data relations into perceived relations in an acoustic signal for the purposes of facilitating communication or interpretation~\cite[p. 5]{Kramer:1999}.\end{quote}
This makes sonification highly suitable for conveying information that changes over time. In the last few years there have been several attempts to develop network sonification systems in order to support network monitoring.  

In order for sonification to serve network monitoring purposes administrators need to have a clear understanding of what is going on in their network environment so they can take appropriate action and prevent malicious activities and misuse of resources. The traffic volumes passing through today's networks are huge which makes it more difficult for them to be represented visually. However, if we enable people to sense and interact with the cyber environment and let the human brain do part of the processing work and to adjust the sound generated to ease analysis this may allow more about the cyber environment to be learned. 
 
\subsection{Situational awareness (SA)}
\label{sec:SA}

Endsley defined situational awareness (SA) is defined as: `the perception of elements in the environment within a volume of time and space, the comprehension of their meaning, and the projection of their status in the near future'~\cite[p. 36]{Endsley:1995}. Because it exists within computers and communication networks the cyber environment severely constrains human perception and so we are reliant on tools to provide perceptual access to what is happening within the network. Vickers et al. described the
situation thus:

\begin{quote}
Many tools on which we rely for situational awareness are focused on specific detail. The peripheral vision (based on a range of senses) on which our instinctive threat models are based is very narrow when canalised by the tools we use to monitor the network environment. The majority of these tools use primarily visual cues (with the exception of alarms) to communicate situational awareness to operators. Put simply, situational awareness is the means by which protagonists in a particular environment perceive what is going on around them (including hostile, friendly, and environmental events), and understand the implications of these events in sufficient time to take appropriate action~\cite[p. 13]{Vickers:2017a}.
\end{quote}

Boyd's OODA (observe, orient, decide, act) loop theory~\cite{Angerman:2004} has added more depth to the understanding of situational awareness. Boyd's theory is based on his study of the decision making of combat pilots and the first stage (observe) involves taking in information about features of the environment. The orientation stage directs attention towards an adversary. The next stage involves deciding what action to take which is followed by acting upon that decision.   

No system can implement the best security measures without interaction with people,  
but it is difficult to maintain high SA levels~\cite{Lakkaraju:2004}. The real-time monitoring of the end-to-end flows and connections in a network is vital to allow better observation and orientation for faster decisions and actions so as to maintain healthy network resources in the face of constant changes in attack methods, motives and behaviours. In general, this work requires high experience and intelligence. Humans are by nature good listeners and are capable of processing auditory events through their experience and intelligence which makes them capable of using sonification for maintaining SA of sensitive cyber environments.  

\section{Sonification and computer network monitoring}
\label{sec:sound}

Sonification is introduced to enable a listener to recognise changes in activities and patterns to enhance comprehension and projection as part of the SA process.   
Sound allows a network administrator to continue monitoring the network while performing other tasks~\cite{Vickers:2011} which may, in turn, decrease frustration and visual fatigue rates.  The concept of changed network behaviour as an indicator of unhealthy activity or intrusion attempts is a reasonable motive for using sonification~\cite{Ballora:2011}. Sonification can have advantages over visualisation in different sectors. For example, real-time sonification using parameter mapping methods is used in the health sector. A recent study showed positive results and a high potential for using real-time auditory feedback-oriented training devices for fitness training or physical rehabilitation to increase the awareness of physiological responses~\cite{Yang:2015}.  

There is a continuing threat of intrusion, denial of service attacks, or numerous other abuses of network resources which requires the monitoring of traffic flows passing through a network ~\cite{DiPietro:2008}. The size of modern network traffic volumes makes it much harder to present real-time information visually~\cite{Fairfax:2014}.  However, there is no clear consensus yet about the pattern of cyber-attacks~\cite{Jajodia:2010}. The assumption is that these behaviours and the rhythm associated with each type of attack should sound different or at least provide an indication of some features of any attack.

Worrall~\cite{Worrall:2015} has described the NetSon project from its exploratory stage to a real-time sonification of network metadata. This project used the information extracted from data volumes by employing sampling techniques to extract a small group of data packets using the sFlow tool~\cite{InMon}. This method provides information about the network's traffic flow rate by making a sonification of sFlow packet data of the traffic such as from printers and servers and load balancing traffic. NetSon also provides information to identify internal and external IP addresses. This tool could be used to support network traffic measurement tools or to identify and classify IP addresses for security purposes.

Mancuso et al.~\cite{Mancuso:2015}  used sonification to help `cyber defenders' to detect evidence of cyber attacks by  using data collected by Wireshark. The data was used offline and the source and destination IP addresses  were sonified using pairs of sequential musical notes separated by 100 ms, while packet size was used to control the loudness of the sound. An experiment revealed no improvement in operator performance when using the sonifications. However, it could be argued that sonification should be tailored so that traffic with specific signatures should sound different from other normal packets, or even that sound should be generated only for malicious signatures. This might increase performance and decrease the stress of the operator.  

Vickers et al.~\cite{Fairfax:2014,Vickers:2017a,nuson-SOCS} applied sonification to the inherent self-organised criticality observed in network traffic. Standard packet capture tools were used to gather network traffic which was then passed to the \textsc{socs} (self-organised criticality sonification) system which sonified the log returns of packet sizes at regular user-specified intervals.  The extracted log returns provide information about the behaviour changes in the network. Knowledge of this behaviour could be used to detect unwanted behaviour. This system has potential to support both network traffic measurement and intrusion detection tools.
 
Wolf and Fiebrink~\cite{Wolf:2013} developed SonNet, a programming interface for sonifying computer network data. The prime motivation behind SonNet was to lower the practical barriers for artists and sound designers interested in accessing network data to create music. SonNet involves packet sniffing and offers network state analysis and easy access to computer network data for composers. The tool supports the sonification of data using the UDP and TCP protocols. SonNet extracts network data at various levels from packet level information to network state information. Level 1 contains information about a single packet, level 2 contains information generated by computing and analysing the single packet information, and level 3 contains information about multiple packets. In a similar vein, Rutz et al.~\cite{Rutz:2015} introduced the SysSon platform for developing sonification applications for different types of users from domain scientists to sonification researchers, composers and sound artists. 

InteNtion (Interactive Network Sonification)~\cite{Giot:2012} is a project targeted at mapping network traffic activities to a musical aesthetic. Network traffic data is converted into MIDI messages and then sent to dedicated synthesisers to generate a dynamic mix of sounds as an interactive soundscape. The system uses IP internet protocols including TCP/UDP segments, using very low-level packet information such as packet size, source and destination IP addresses and type of service. The work is still considered to be experimental and the system needs more development and better mapping to support network traffic monitoring. However, it provides an innovative way to monitor a network by using the entire data flow to create music.  

Earlier work done by Ballora and Hall~\cite{Ballora:2010} explored the detection of intrusion signatures and patterns using human aural and visual recognition abilities to detect intrusions in real-time. IP addresses and return codes were used to generate sound as an informative and unobtrusive listening environment to develop web traffic SA. Ballora et al.~\cite{Ballora:2011} conducted another sonification experiment with a computer network based on socket connections using information such as the date and time of exchange and the sender's  and receiver's IP addresses and port numbers. Ballora et al.~\cite{Ballora:2012} have also described the use of sonification in the detection of anomalous events.  Sonification should enable the listener to differentiate between normal and anomalous network behaviour and to develop an understanding of what is actually happening in the network. 

Kimoto and Ohno~\cite{Kimoto:2002} introduced the Stetho network sonification system which was aimed at system administrators.  NetSound  was built on top of Stetho as a tool for end users. Stetho used network traffic information to generate sounds which provide the network administrator with information about the traffic. Stetho reads the tcpdump \cite{tcpdump} commands, then uses them in regular expressions to generate corresponding MIDI events. Stetho processes each packet in the traffic. However, Stetho failed to detect all events and intrusions. Delays in sound generation and poor MIDI messages generated further problems. 

Chafe and Leistikow~\cite{Chafe:2001} developed a tool for the measurement of round trip time when using a sequence of standard ping utility events  to gather information about the quality of service of a network path, such as packet loss. They discussed the need to evaluate paths which carry interactive media streams in collaborative environments. They designed a stream-based method for the direct display of the critical qualities to the ear by continuously driving a bidirectional connection to create sound waves. They changed the network path to acoustic medium which their probe sets into vibration. Temporal levels of musical foreground, middle-ground and background are heard in the melodies generated from correspondence data.

\section{\Son}

The sonification of high-speed computer networks demands both high throughput and flexibility to handle and recognise new threats. It is possible that sonification is a viable solution to this problem and could allow an administrator to listen in real-time to the state of each traffic flow. As a solution to these problems and issues, we propose \Son\ --- Sonification of Networks for SiTuational AwaReness --- to be used by network administrators as a monitoring tool to facilitate the acquisition and maintenance of network situational awareness. \Son\ would assist with the maintenance of security, awareness of anomalous events such as attacks, maintenance of network health through monitoring and tuning, and increasing the understanding of the cyber environment which is vital  for network management the use of diagnosis to support the recognition phase in the situational awareness process.

A traffic flow is a flow of packets in single connection between a source computer and a destination~\cite{Brownlee:1999}. A single flow can be identified within a certain time period by its source and destination IP addresses and ports and its protocol layer (such as TCP, UDP and ICMP). As part of our technical solution, we have created new flow type called  IP flow which is identified within a certain time period by its source and destination IP addresses and protocol only (no port information. 
Thus, \Son\ uses these two flow types (traffic flow and IP flow). 

\Son\ uses events to generate sounds. A flow event is a change in the behaviour or operation of a flow (traffic or IP). A single event represents a combination of features of a flow while a set of events represents flow behaviour which, in turn, represents the state of the network traffic. 
 
In the TCP protocol, the header contains nine control flags, six of which (FIN, SYN, RST, PSH, ACK and URG) are used by \Son. Values of 1 and 0 denote whether a flag is set or unset, and the packet's type is determined by those flags that are set. A packet's type determines its role and function within the network traffic. Therefore, \Son\ collects counts of each packet type for both traffic- and IP-flows. A flow's status is determined by the respective packet type counts. \Son\ allows its user to listen to the status of the flows in traffic by playing sounds that represent the flow behaviours. 

Thus, \Son\ makes information about traffic perceptible  allowing the network administrator to make decisions about network operation on the basis of recognising the sounds that describe the network environment.  \Son\ allows users to set specific sounds for different flow status types and to tune the thresholds for triggering the sounds. What makes \Son\ distinctive compared to other available tools is that it allows the user to monitor general and specific behaviours in a human understandable form.  

When designing a sonification system with the purpose of monitoring a system or network activity to gather administratively useful information, the design will involve a number of conditions and requirements. The sound has to support extended periods of  listening, changes in status have to be easily grasped and accidental events have to be immediately noticed~\cite{Kimoto:2002}. 
\Son\ sonifies each flow in a connection and collects information about the connection state by periodically gathering online flag information from each flow. Traffic features are extracted from the flag information aggregations and \Son\ then represents these features in a soundscape.

The term ``soundscape'' was introduced by Murray Schafer~\cite{Schafer:1977} and describes the sonic properties of landscapes. Sounds are a continuous and active property of all landscapes and soundscape ecology is generated from the sounds and spatial temporal patterns as they occur in a landscape environment, where each sound has special ecological characteristics~\cite{Pijanowski:2011}. In \Son\ the network environment is transformed into an acoustic environment as a soundscape and the combinations of sounds represent the current state of the network, just as the combinations of sounds in a landscape provide information about what is happening in the environment.

\Son\ transforms the network environment to the soundscape of a forest (though it is fully configurable and allows any other soundscape to be used as desired). Just as a person would be able to infer information about what is happening in a forest by the sounds they hear, sounds in the soundscape represent events and unexpected or particularly loud sounds can draw the listener's attention to traffic behaviour that is out of the ordinary.

Using recorded sounds in sonification can be difficult, as there are limitations on how recordings can be used to represent traffic while still sounding realistic~\cite{Kramer:1994}. However, the use of recorded sounds is better than synthesised sounds, because it enables users to link events to familiar and understandable sounds. Sounds from a natural environment such birds tweeting or animal sounds are easier to describe than artificially synthesized tones which may rely on specific terminology such as frequency and timbre~\cite{Wolf:2015}. The sounds provide us with immediate awareness of the types of events that are happening:
\begin{quote}
 Modern cognitive science believes that to be able to read sound in this way, listener must have some inner understanding of how the properties of physical events are reflected in the sounds they make ~\cite[p. xviii]{Kramer:1994a}.
\end{quote}

Therefore, a monitoring operator requires a good understanding of communication protocols and theoretical and practical knowledge about the expected behaviour in computer networks.  \Son\ allows the user to make a relation between the meaning of the recorded sound and the event mapped to within the network environment. For example, a forest on a normal day will produce sounds of typical birds and animals, perhaps with a very light breeze. These events can be used to describe the normal state of a computer network environment. While human-made sounds, the sounds of predators and changes in weather (such as rain and thunder) can be used to represent abnormal or malicious network activity. \Son\ sonifications are generated using an event mapping method based on flag state information collected from each TCP packet for each flow in the network. This specialised abstraction of network features is extracted from the raw flow packets and transformed into classified sound groups of natural and human-made sounds. 

\section{\Son\ and network traffic monitoring}
\label{sec:ma}

Commonly, administrators try to look directly at network traffic to understand it using tools such as Wireshark \cite{Wireshark}. Network traffic volumes can be huge and the majority of the traffic involves normal data packets travelling between legitimate users on the network or across the internet. TCP/IP packets carry control flags to allow the data to be received in sequence and to protect it from loss. In TCP/IP, if receipt of any packet is not confirmed by the destination, it will be sent again. In contrast, in the UDP protocol any packet sent will be considered as received and packets will be processed in the order they arrive regardless of whether the routing has caused them to be received out of sequence.  In TCP/IP approximately 30\%-40\% of traffic concerns packets which are very important to administrators for enabling them to understand immediately what is happening in their network environment~\cite{Shah:2004}.   
This means that the TCP/IP control packets SYN, SYN ACK, ACK, FIN and RST provide most of the information about network traffic state. UDP packets have to be monitored in such a way that allows administrators to recognise the current state. TCP/IP traffic represents more than 85\% of packets entering and leaving a system or computer network~\cite{Shah:2004,Balram:2008}; therefore, TCP traffic is considered a priority.

Network administrators typically identify anomalies in traffic from two sources. The first is simple network management protocol (SNMP) data from queries to network nodes. However, the data collected from the SNMP management information base (MIB) is wide ranging, and contains activity statistics such as total packets transmitted at a node. This source can only provide statistics about volumes of packets and bytes which provide useful information but cannot be used to understand the behaviour in the traffic flows and connections in the network. The second source is  the monitoring of end-to-end packets, flows or connections. This data contains protocol-level information. This second source is typically used by intrusion detection systems. These two sources offer a practical base for the identification and recognition of anomalies as part of situational awareness~\cite{Barford:2002}. 

\Son\ uses the second source and collects data by sniffing the traffic passing through a switch or a router from the mirroring outlet in real time or by reading stored PCAP files captured by any other available packet sniffing programs. The sniffer act as a sensor that collects traffic information periodically.

Several types of monitoring systems use network usage patterns for detection, measuring usage and summarising usage statistics based on user-defined parameters, and contrasting measurement aggregates with predefined thresholds and then responding when thresholds are met or exceeded or following queries from a security analyst~\cite{Bowman:1997}. 

Most detection methods (especially IDS)  depend on packet headers or the payload or a combination of both to detect attacks and malicious activity. In anomaly-based systems analysis of the packet's payload is used to differentiate between normal traffic and anomalous activity. Signature-based systems rely on matching patterns with a database of the signatures of known attacks. The advantage of anomaly-based systems is that unlike signature-based systems, they can detect attacks without any delay since new attacks can be detected as soon as they happen, while signature-based systems cannot detect novel attacks and can only match against known attack signatures~\cite{DiPietro:2008,Srinivasan:2006}. While anomaly-based systems can detect novel attacks they generate more false positive results and so risk blocking legitimate activity. Identifying the state of traffic from encrypted applications is a critical issue for numerous network tasks. In-depth packet inspection requires decryption in most cases, and this would affect any detection mechanism especially when trying to operate  in real time~\cite{Alshammari:2011,Stallings:2014,Bernaille:2007}. 

\Son\ uses packet header information to generate sounds which periodically represent the status of aggregated packet information for multiple flows in the network. It is an anomaly-based system which generates different sounds according to the network state. This method can be used to provide a general or specific sonic representation of the traffic behaviour. Any changes in sound combinations then represent a new state or behaviour. An advantage of this approach is that an administrator using \Son\ can interact with the system and change and create the features to be sonified and assign sounds to those features. \Son\ is an additional tool that enables administrators to discover changes in and learn more about their environment in a way that enables the human mind to comprehend the mechanism of these changes and their causes.

A  security system using real-time monitoring for situational awareness has to show changes in flow and connection states as they happen and provide an indication to the administrator about immediate events. \Son\ targets this type of monitoring to support existing security tools, acting as an additional tool aimed at raising situational awareness levels.

\section{\Son\ design}

Computer network defence requires analysts to detect both known and novel forms of attack in massive volumes of network data. Visualisation tools would potentially assist in the discovery of suspicious patterns of network activity, but few analysts so far are leveraging sonification technologies in their current security practice. We have designed \Son\ to suit the work practices and operational environments of those analysts.

The work's novelty consists in sonifying in real-time the extracted features of network traffic based on the control flag status of the packet header and the techniques developed to handle the interaction of the user with the system with the aim of increasing situational awareness levels. \Son\ inspects the flag statuses of each packet in the flows and extracts features by  periodically counting each packet type and counting the number of flows and then uses this information to control the resulting soundscape. This results in a system that is complementary to and more informative than visualisation methods, but which can provide only limited goal-oriented information. This type of sonification, which allows the representation of large traffic volumes by representing traffic flow and IP flow states to reduce the amount of traffic information presented to the user, has not been done before.

In many systems, changes in performance could be used to indicate the vulnerability or robustness of a computer network~\cite{Criado:2005}. Equally, changes of sounds could be used to indicate changes in network behaviour.  The first goal of the design of the sonification system as part of the situational awareness process is either to monitor network assets or the network gateway and to find a way to sonify network component activity and traffic behaviour to enable the listener to detect any misuse or anomalous behaviour. This anomaly detection approach must first determine the normal behaviour of the object being monitored, and then use deviations from this baseline to build experience and knowledge to detect and identify possible malicious activities. 

Monitoring tools try to present administrators with a complete representation of their complex network. Better network monitoring tools should allow administrators to perceive changes in their network in order to allow them to react immediately, and learn and understand more about the cyber environment. A real-time sonification monitoring tool should be able to do or assist with the following:

 \begin{itemize}
 \item \textbf{Identify and recognise malicious traffic:} Malicious traffic such as probes and denial of service attacks should be indicated.
  \item \textbf{Provide information about incidents or changes in behaviour:}  An incident or change in traffic behaviour should be reported to allow the user to recognise which flows are malicious.
 \item \textbf{Represent network behaviour sonically and in a non-fatiguing and non-annoying way:} Sounds representing states have to be easily recognised and linked together by the user to allow comprehension as part of the situational awareness process. 
\item \textbf{Offer practicality:}  Use of the system should be convenient for both incident response and real-time monitoring.
 \item \textbf{Indicate compromised machines:} A machine compromised  by a hacker or malicious software such as worms or viruses should be indicated when ever possible.
 \item \textbf{Offer high throughput and flexibility:} The system should be able to handle large amounts of data in a timely manner and its operation should not be CPU-intensive.
 \end{itemize}

\subsection{Monitoring requirement of the tool}

The output of such a system is meant to help the user to identify changes in traffic behaviour or recognise attacks immediately as part of the situational awareness process. This awareness is important and its lack could be costly and decisive for an organisation. It is important that the monitoring tool assists the user to analyse and interpret the traffic in the correct manner. Various common requirements for forensics analysis, visualisation and sonification tools for monitoring are given in the literature  ~\cite{Carrier:2003,Malandrino:2003,Neuhoff:2002,Bass:2000,Baier:2006} and the following requirements are based on them:

 \begin{itemize}
 \item \textbf{Usability:} Data sonified at the lowest packet information level would result in huge volumes of information which would be to difficult for the user to interpret. Therefore, the representation of this information by sound has to be designed so that the user can recognise normal and malicious activities. The information has to be represented by distinct sounds so that it is not misinterpreted.
\item \textbf{Cognitive processes:} The time it takes to learn how to use and understand the system should be minimised.
\item \textbf{Comprehensive:} The sounds generated have to represent, as far as possible, all output data at a given level of abstraction. 
\item \textbf{Accuracy:} The tool should guarantee that the output sounds are clearly distinguishable and that the margin of similarity should be presented to the user, for example as a log file, so that it can be confirmed and interpreted correctly.
 \item \textbf{Deterministic:} The tool should always generate the same output sounds when presented with the same input dataset or traffic when using same sound design.
 \item \textbf{Verifiable:} To ensure the accuracy of the tool, it should be possible to verify the results. This could be done manually or by using another tool.
 \end{itemize}

\subsection{Design solution}

This section considers a design that can fulfil the requirements of real-time monitoring for situational awareness. The practical issues associated with the selected design are also discussed.

\begin{figure}[!h]
\includegraphics[scale=0.4]{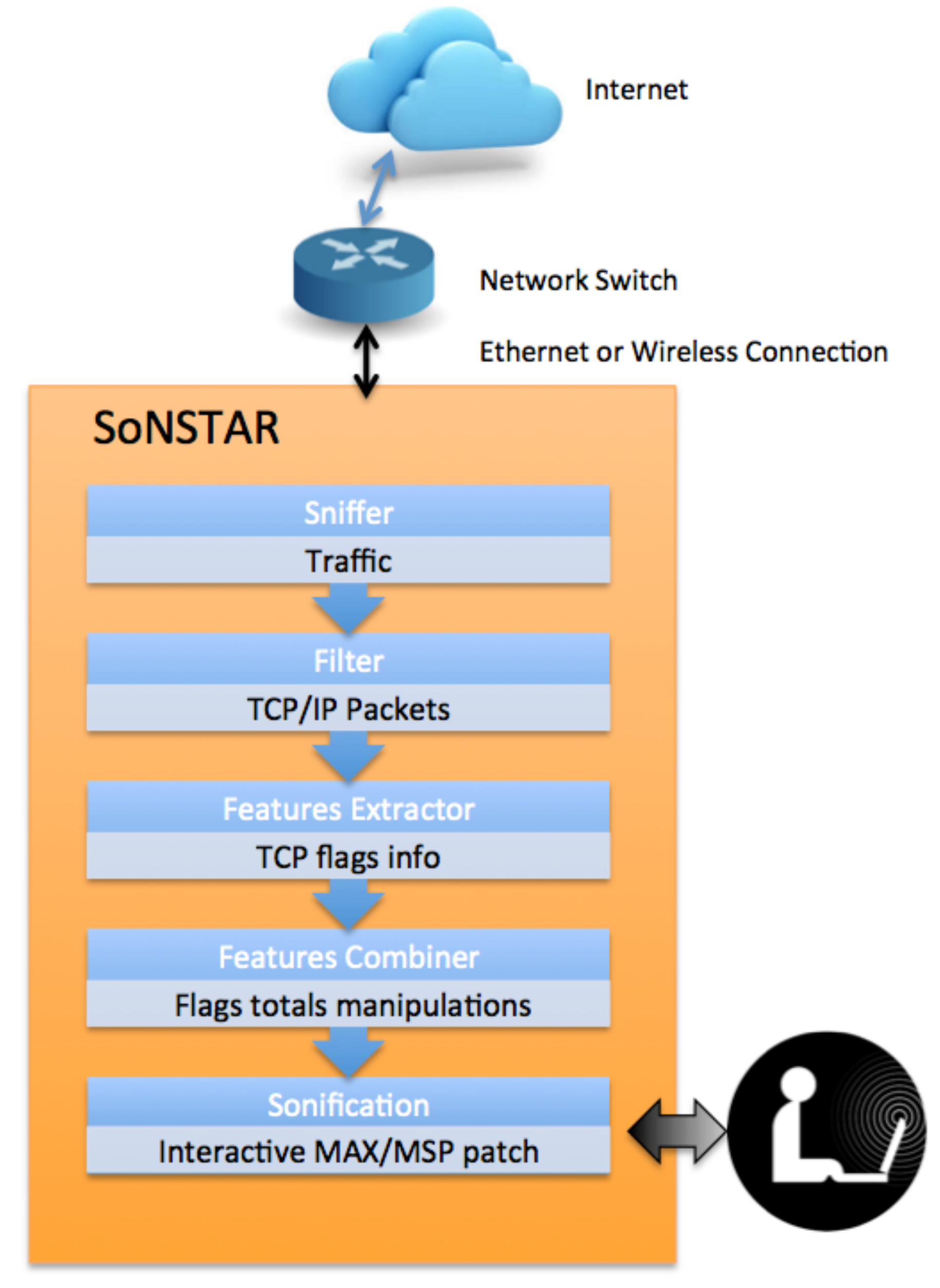}
\caption{{\bf SoNSTAR architecture}
 The major components of the system.}
\label{fig1}
\end{figure}

The \Son\ architecture diagram is illustrated in Fig \ref{fig1}. The system is implemented in Python using the pcapy and dpkt libraries and Max/MSP. The Python engine captures and processes the packet information and passes data to the Max/MSP patch which generates the audio. 

\begin{figure}[!h]
\includegraphics[width=\linewidth]{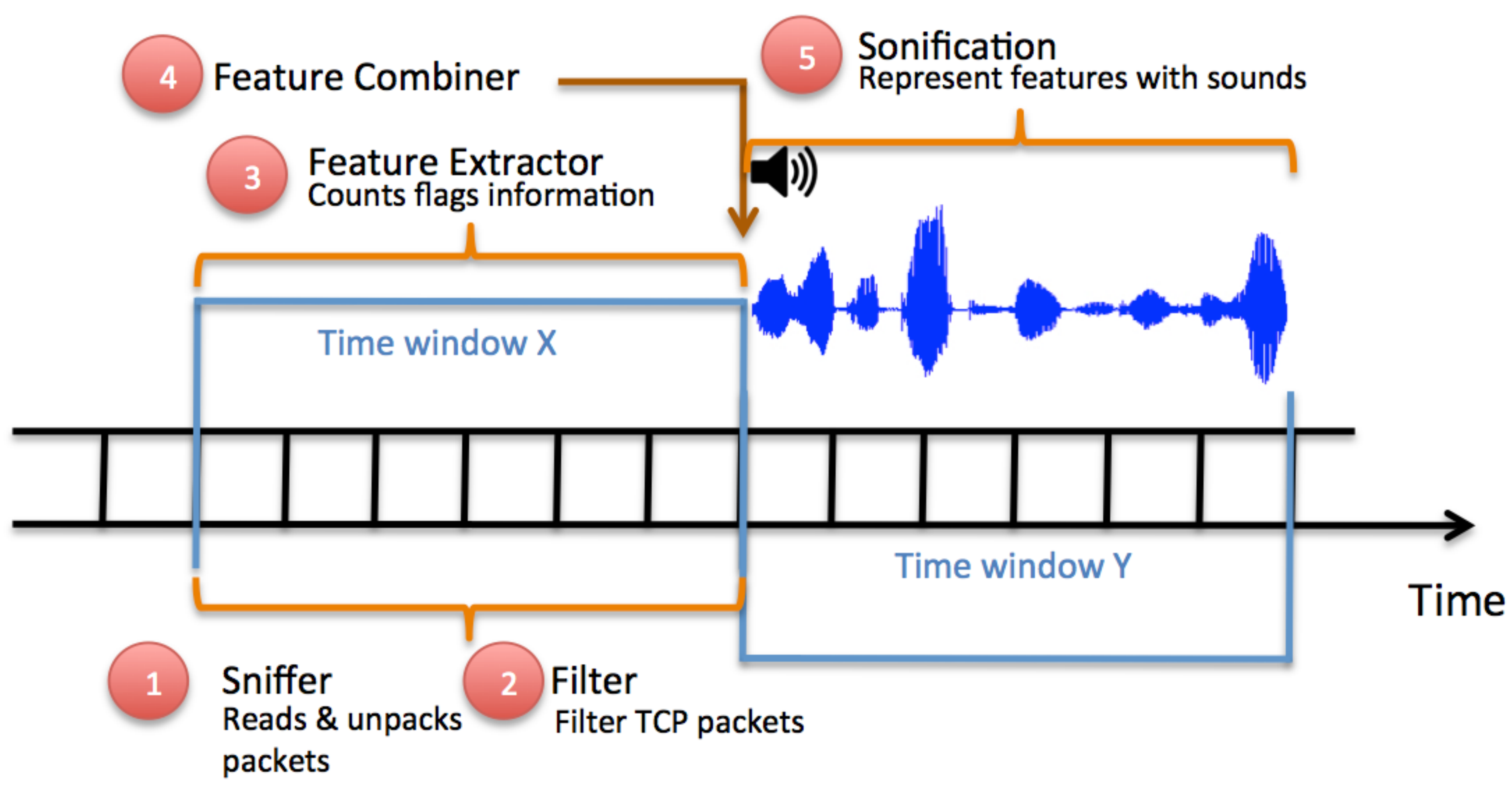}
\caption{{\bf Time window processes}
  \Son\ aggregates flow data across time windows.  This figure shows the process timing and sequencing across two time windows, $X$ and $Y$.}
\label{fig2}
\end{figure}

\Son\ uses time a window period to arrange and control the timing of the operation of each process within the system (see Fig \ref{fig2}). \Son\ reads packets and unpacks them and filters the TCP packets and extracts counts during time window \textit{X}. At the end of each time window, features are combined to generate higher-level aggregate features. The selected features are then represented as recorded sounds. These sounds are played during the next time window \textit{Y}. 

The main \Son\ algorithm is shown in Algorithm \ref{algorithm}. The \Son\ system comprises five blocks described below.

\begin{algorithm}
\small
\caption{\Son's main algorithm}
\label{algorithm}
\begin{algorithmic}
\State\textbf{Set Timewindow period}
\State Sniff packet and Get start time
\If{$Packet == arrived$}
	\State Unpack ethernet header
	\State Extract EtherType
	\If{$EtherType == \texttt{0x0800 or 0x86DD}$}\Comment{IP packet}
		\State Unpack IP header
		\State Extract source and destination addresses
		\State Extract transmission protocol
	\Else
		\State Get next packet from the sniffer
	\EndIf
	\If{$Protocol number == 6$}\Comment{TCP packet}
		\State Unpack TCP header
		\State Extract flags information according to incoming or outgoing
		\State Count flags status according to incoming or outgoing
		\If{$Timewindow period== finished$}
			\State Extract current flag's features 
			\State Extract new features from Features Combiner
			\State Apply thresholds to selected features
			\State Send messages to Max/MSP for sonification
		\EndIf
		\State Get next packet from the sniffer a new Timewindow started
	\Else
		\State Get next packet from the sniffer
	\EndIf
\Else
	\State Get next packet from the sniffer
\EndIf

\State\textbf{Max/MSP Patch}
\If{$messages == arrived$}
	\State Play sound of similar messages once
\EndIf
\end{algorithmic}
\end{algorithm}

\subsubsection{Sniffer}
The main input to the system is  the raw traffic packets passing (incoming and outgoing) through the network and the Sniffer reads these packets in real time. 
 
\subsubsection{Filter}
The Filter unpacks each ethernet frame, extracting the packet header information, and sending only TCP packets to the Feature Extractor. A TCP/IP packet has an EtherType value  of \texttt{0x0800} or \texttt{0x86DD} (denotes IP protocol) and a transmission protocol number of 6. 

\subsubsection{Feature Extractor}
 
Next, the Feature Extractor picks up each TCP packet, checks the flag values,  and determines the packet type. If this flow has not been seen before it creates a new Traffic flow and IP flow and sets the counter for the current packet type to 1 for each flow. If the flow already exists the feature extractor increments its packet type counts by 1 according to the packet's direction (incoming or outgoing). This update happens for both flow types (traffic flow and IP flow). At the end of the time window, the set of traffic flows with their packet type counts and the set of IP flows with their packet type counts, in addition to number of traffic flows and number of IP flows are passed to next stage. The full feature sets for both flow and IP-flow packets can be found in the supplemental material (see \texttt{S1\_Appendix} in the supplemental material). 

At the end of each time window \Son\ creates two logs file reports consisting of the entire traffic flows and IP flows with their packet type counts respectively for any post-hoc inspection and review that may be required.
 
\subsubsection{Feature combiner}

The Feature Combiner enables the user to create new features by adding or subtracting particular flags (see Table \ref{table1} for some examples). This enables the user to target specific flow events. Some of these combinations could be set according to user needs and understanding of the TCP protocol behaviours and rules. Some could be built over time while listening to and learning about the network environment's  behaviours and sounds. 
 
For example, TCP requires the use of specific mechanisms to establish connections between source and destination hosts. An established process is called the three-way handshake. The first step in the process is that the source (S) sends to the destination (D)  a TCP packet with the SYN flag set. Next, D replies to S with a packet with the SYN and ACK flags set to  acknowledge and accept the connection. Finally,  S sends a packet to D with the ACK flag set indicating acknowledgment of the agreement. In this way the handshake process is successfully completed and the connection is established. After the exchange of data and at the end of the connection, either side will terminate the connection by sending a TCP packet with the FIN flag set~\cite{Schuba:1997}. Therefore, each flag's status gives us information about the flow and changes in flag status represent what is happening in the network.

At this stage of \Son\ design we have created some new features from  previous IP flow features (see \texttt{S1\_Appendix}) provided by the Feature Extractor (see Table~\ref{table1}).  All of these features are now available for sonification.

\begin{table}[!ht]
\scriptsize
\centering
\caption{Feature combinations}
\begin{tabular}{L{1.6cm}L{4.5cm}L{1.5cm}}
\toprule
\textbf{Feature Combination} & \textbf{Definition} &\textbf{Normal range} \\ 
\midrule

FC 1 & SYN-out-IP $-$ SYN-ACK-in-IP & $\leqslant 4$  \\

FC 2 &  SYN-in-IP $-$ SYN-ACK-out-IP & $\leqslant 4$  \\

FC 3 &  FIN-out-IP $-$ FIN-in-IP & $\leqslant 9$ \\

FC 4 &  FIN-in-IP $-$ FIN-out-IP  & $\leqslant 9$ \\

FC 5 & SYN-in-IP $+$ SYN-out-IP $-$ FIN-out-IP & $\geqslant$ RST-out-IP  \\  

FC 6 &  SYN-in-IP $+$ SYN-out-IP $-$ FIN-in-IP & $\geqslant$ RST-in-IP \\

FC 7 & FIN-in-IP $-$ FIN-out-IP $-$ RST-out-IP  & $\leqslant 9$ \\

FC 8 & FIN-out-IP $-$ FIN-in-IP $-$ RST-in-IP & $\leqslant 9$  \\ 
\bottomrule
\end{tabular}
\begin{flushleft} Illustration of the way packet counts (by flag type) are combined to denote specific feature combinations.
\end{flushleft}
\label{table1}
\end{table}

\subsubsection{Sonification}
 
The final block in the system is Sonification. To make sense of the sonification we have to assign sounds according to event conditions and thresholds and according to the understanding of flag status mechanisms for both flow types. Knowledge of these events could be learned over time while listening to the network environment, tuning the thresholds and experimenting with conditions to target particular behaviours and exploring log files. 

Through development of this design recorded natural sounds have been assigned to various features to create a network soundscape environment. 
By operating \Son\ and listening to sounds and manipulating event conditions and tuning thresholds, new events and feature combinations can be defined (such as those new features listed in Table \ref{table1}). Threshold values could vary according to the characteristics of the network being monitored

Of the many features that could be monitored for intrusion detection purposes, some are truly useful and some are less significant, and may indeed be useless. A standalone IDS might generate many false positives or could ignore an anomaly (false negative) depending on its settings. There is no clear analytical model that provides the basis for a mathematical formula to precisely describe the input-output relationship~\cite{Mukkamala:2005}. Therefore, using \Son\ would provide that missing understanding of the decisions made by an IDS and allow its user to gain knowledge through monitoring  the real behaviour and events of the flows within the traffic.  

Every network is a unique environment. Relationships between features are important when applying sounds to the events chosen. This is what gives \Son\ a real advantage in exploring a network environment because the understanding of the traffic environment can be improved by taking into account feature relations. 
The idea behind using different recorded sounds from nature and human-made sounds to represent the network environment is to transform the experience into an interactive soundscape environment. The sounds generated express the behaviour of flows and their deviations from the normal state in order to increase situational awareness.

\subsection{Representational techniques}

\begin{figure}[!h]
\includegraphics[width=\linewidth]{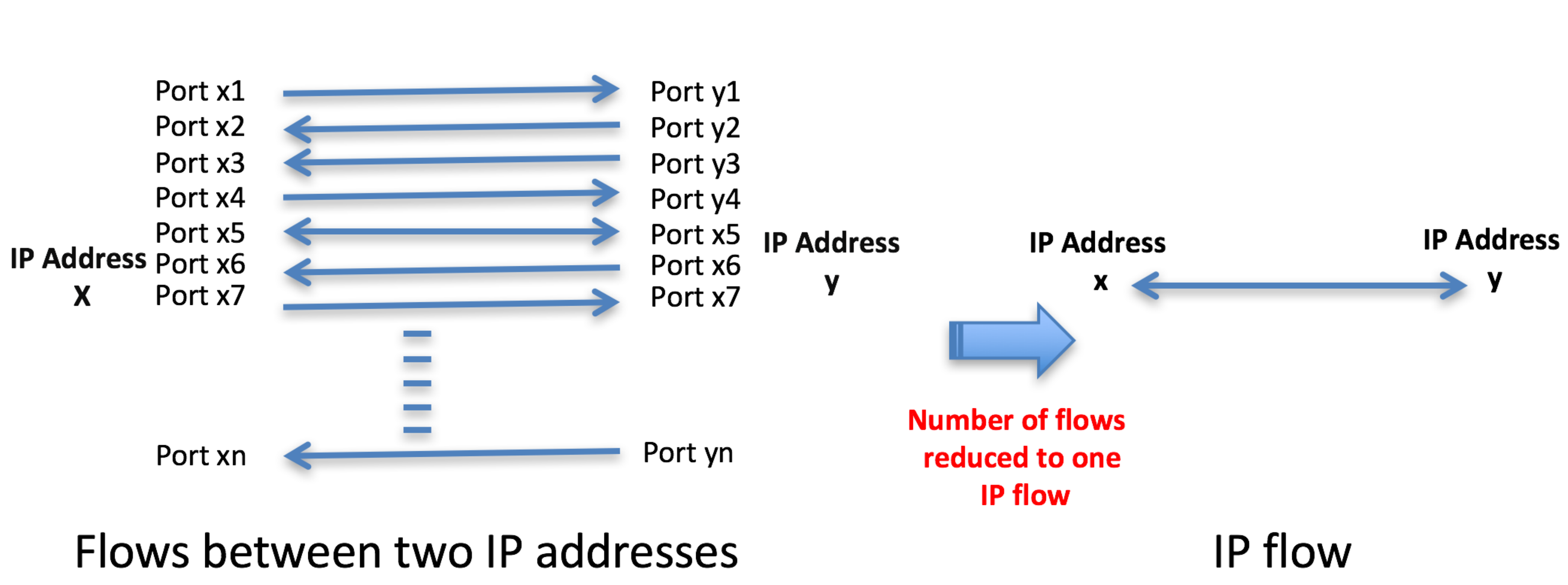}
\caption{{\bf Conflation of multiple traffic flows to one IP flow}
Seven traffic flows between different ports on the same sending and receiving hosts are reduced to a single IP flow.}
\label{fig3}
\end{figure}

Sonic representation is a challenge because of the huge volumes of traffic passing through each connection in the network. Each connection has a high potential number of flows depending on the nature of that connection and its purpose. \Son\ reduces the complexity of representing huge volumes of traffic by two methods. The first considers IP flows rather than traffic flows. A number of traffic flows could exist between any two hosts as each traffic flow is specific to a single port number. IP flows are not concerned with port numbers so the number of flows between any two hosts is reduced to one for sonification purposes (see Fig \ref{fig3}). In the second method \Son\ maintains counts of the packet types for each traffic flow to update the soundscape at a user-specified interval. Since network traffic consists of a number of flows which can be similar in their condition, so similar flows can be expressed once so that there is no repetition of the same sound. By doing this we have reduced the number of flow events that need to be sonified.

Recorded sounds (such as birds or rain) represent discrete events by playing a single natural sound every time the event occurs. The sounds chosen are diverse in nature and easily distinguishable by the listener. 

\subsubsection{Tuning the system}
\begin{figure}[!h]
\includegraphics[width=\linewidth]{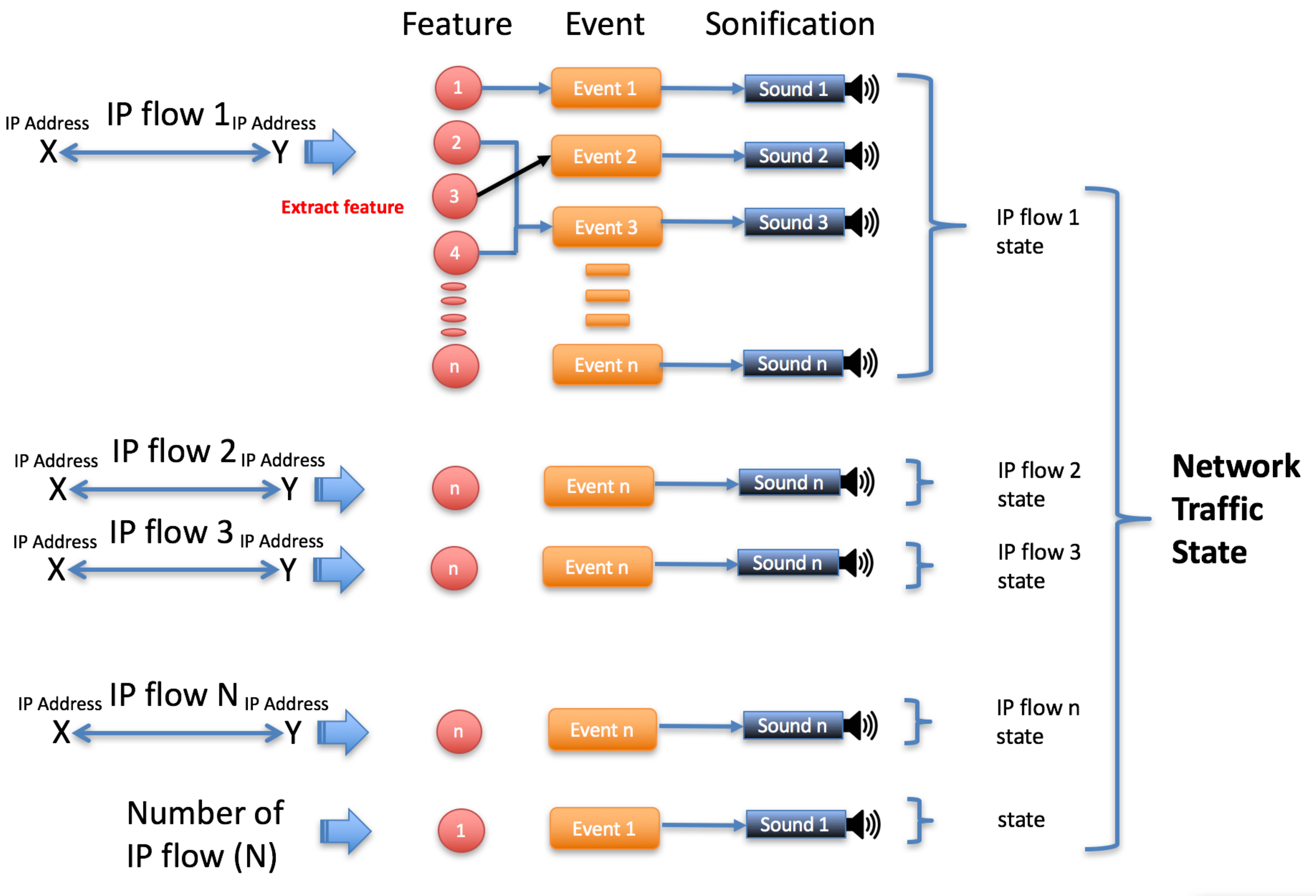}
\caption{{\bf IP flow representation}
Illustration of multiple IP flows containing a range of different events and even combinations are mapped to different sounds resulting in a sonic representation of the overall traffic state.}
\label{fig4}
\end{figure}

One begins to tune \Son\ for a particular network by starting with the three-way handshake mechanism and assigning it to a chosen sound. Then,  each flow event of interest is mapped to a sound and then its frequency of occurrence is listened to over time in order to get a sense of its impact on network behaviour. The event's feature threshold value can then be adjusted to suit. It was noted during development that certain events tend to occur normally in every network or dataset. Network mechanisms and activities which are confirmed as normal events were mapped to sounds from a forest birds collection. Forest birds were used because they represent the normal state of a forest. Sounds that do not belong to the normal state of a forest were then used to represent rarer, unusual, or anomalous events.  Fig \ref{fig4} shows an example sonification of IP flows to represent network traffic state. (Listen to the file \texttt{S1-Audio} of normal traffic sonification in the supplemental material).

Events which are outside the normal range are represented according to the main flag type that caused that event. Sound representation is divided into five categories. The first category of network states represents ongoing events related to SYN or SYN-ACK packets (or combinations thereof) and is represented by weather-related sounds of rain or water.  For example, the soundscape changes from rain to heavy rain to rain and thunder according to the number of packets that caused the event. 

\begin{figure}[!h]
\includegraphics[width=\linewidth]{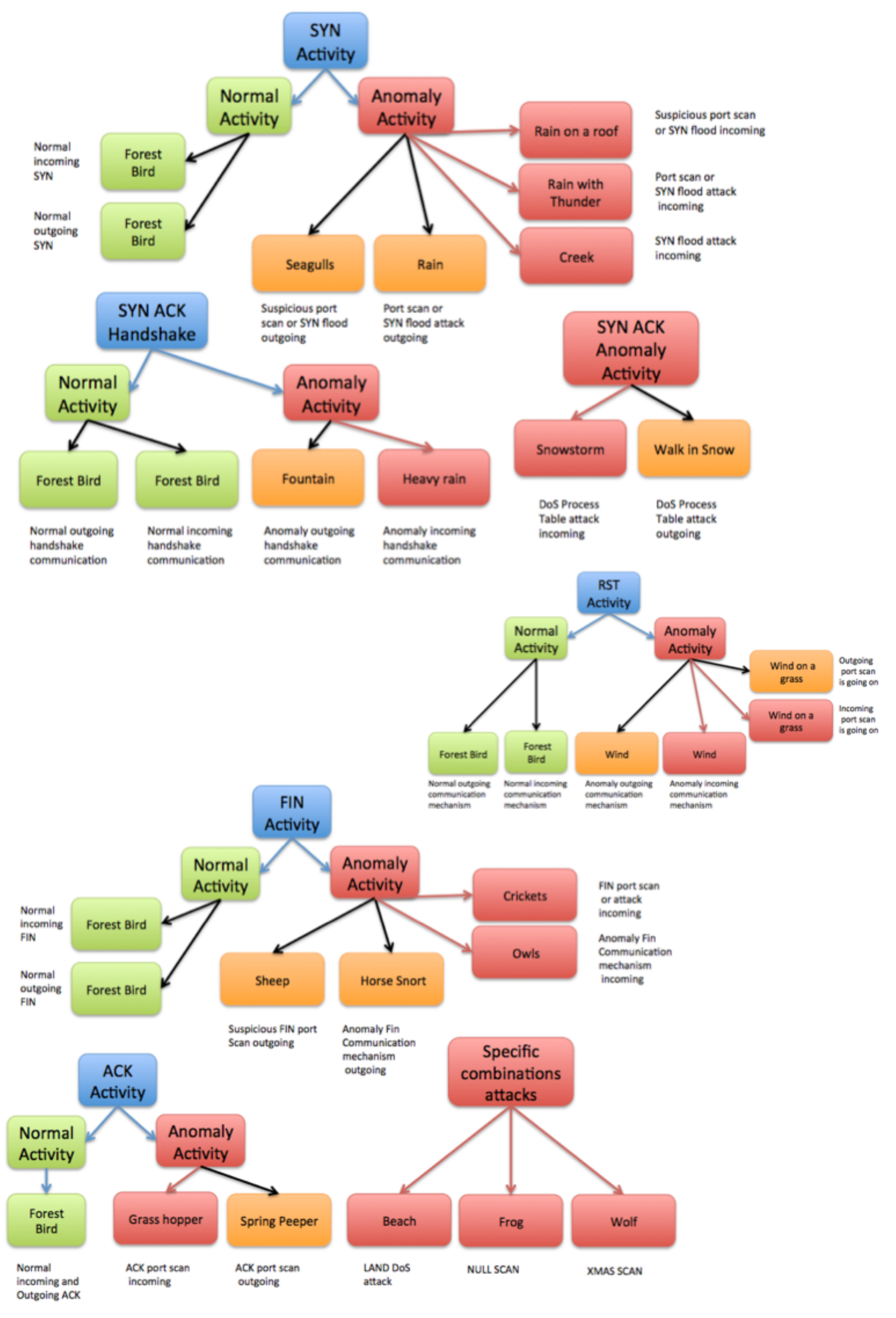}
\caption{{\bf Event representation}
Illustration of different events (identified the main flag type) being mapped to discrete sounds the \Son\ soundscape.}
\label{fig5}
\end{figure}

The second category represents ongoing FIN, ACK, URG, PSH or NULL packets (or combinations thereof) and is represented by animals or unusual birds. The third category represents ongoing RST events and is mapped to wind sounds. For example, when any host sends a high number of RST packets the sonification reflects the change in network state by playing a wind on grass sound; if the RST packet changes usual behaviour in relation to SYN and FIN packets, a heavy wind sound is played. The fourth category  represents ongoing events  related to traffic- or IP-flow counters  and is represented by sounds of fire in the woods. The fifth category represents ongoing events confirmed as normal conditions and is represented by usual forest birds forming ongoing background sounds. Fig \ref{fig5} shows an example of event representation in \Son. 

For a better representation, incoming and outgoing events of the same type are represented such that incoming events are given more worrying and louder sounds (more dangerous- or urgent-sounding versions of the sounds) than outgoing events  which are quieter which are mapped to non-alarming animal sounds. Furthermore, it was observed that several events tend to occur together or in specific sequences for particular types of attack. Therefore, their sequenced sounds were example of behaviours that were learned as \Son was used to begin exploring network traffic. It is posited that the information about network traffic provided by \Son\ can assist with the recognition of anomalies, both of known and unknown (not previously encountered) types.

Sound design and representation depend very much on personal taste and targeted behaviour. \Son\ provides the user with a choice of sound sets (e.g., forest, weather, and animals sounds, or even human-made ones) and assigns sounds according to the event features the user wishes to monitor. 

 \subsection{\Son\ features sound mappings}
 
The features used for sonification are aggregation counts of the flag status of each flow type in the traffic. For each feature thresholds are set such that sounds are generated only when the counts exceed the threshold. Users can select the thresholds appropriate to their network environment. A set of default mappings was created based on an understanding of TCP protocol theory and running \Son\ several times whilst carrying out simulated attacks in order to learn about traffic features. The thresholds used do not represent \emph{a priori} fixed rules. However, experimenting with these thresholds requires an understanding of the flag relations in the TCP protocol. Network traffic is not static and what can be normal traffic behaviour in one context could be malicious elsewhere, and thus the expected numbers of flows could vary depending on the purpose of the network. The default event-to-sound mappings are listed in Table~\ref{table2}.

\begin{table*}[!ht]

\centering
\caption{Feature-to-sound mappings.}
\begin{tabular}{cL{13.6cm}c}
\toprule
\textbf{No} & \textbf{Feature Conditions} & \textbf{Sound }  \\
\midrule

1 & SYN-in-IP \textless 30 and SYN-ACK-out-IP \textgreater 0 and ACK-in-IP \textgreater 0 and RST-out-IP \textless 10 & Forest bird\\

2 & SYN-in-IP \textgreater 10 and SYN-in-IP \textless 30 and PSH-ACK-out-IP \textless 6 & Rain on roof \\

3 & SYN-in-IP \textgreater 20 and SYN-ACK-out-IP \textless 10  & Rain on roof \\

4 & SYN-in-IP \textgreater 300 and SYN-ACK-out-IP \textless 50 and SYN-in-IP \textless 1000  &  Thunder \\

5 & SYN-in-IP \textgreater 1000   &  Creek \\

6 & SYN-out-IP \textless 10 and SYN-ACK-in-IP \textless 2 and ACK-out-IP \textless 3  & Rain \\

7 & SYN-out-IP \textless 30 and SYN-ACK-in-IP \textgreater 0 and ACK-out-IP \textgreater 0 and RST-in-IP \textless 10 & Forest bird \\

8 & ACK-in-IP \textgreater 1 and the rest of IP flow feature equal 0 & Seagulls \\

9 & ACK-out-IP \textgreater 1 and the rest of IP flow feature equal 0 & Loon \\

10 & FIN-in-IP \textgreater 9 and FIN-in-IP  \textgreater SYN-out-IP and FIN-in-IP \textgreater SYN-in-IP and FC-4 \textgreater 10 & Cricket \\

11 & FIN-in-IP \textless 50 and (FIN-in-IP  \textless = SYN-out-IP or FIN-in-IP \textless = SYN-in-IP ) & Forest bird \\

12 & FIN-out-IP \textgreater 9 and FIN-out-IP  \textgreater SYN-out-IP and FIN-out-IP \textgreater SYN-in-IP and FC-3 \textgreater 10 & Sheep \\

13 & FC-7 \textgreater 9  & Owl \\

14 & FC-7 \textless 10  & Forest bird \\
 
15 & FC-8 \textgreater 9  & Horse snort \\

16 & FC-8 \textless 10  & Forest bird  \\

17 & NULL-in-IP \textgreater 0  & Frog \\

18 &  NULL-out-IP \textgreater 0  & Frog \\

19 & URG-PSH-FIN-in-IP \textgreater 0  & Wolf \\

20 & URG-PSH-FIN-out-IP \textgreater 0  & Wolf \\

21 & LAND-in-IP \textgreater 0  & Beach \\

22 & LAND-out-IP \textgreater 0  & Beach \\

23 & RST-in-IP \textgreater 25 and ACK-in-IP \textless 250   & Wind on grass \\

24 & RST-out-IP \textgreater 25 and ACK-out-IP \textless 250   & Wind on grass \\

25 & FC-1 \textgreater 4  & Fountain \\

26 & FC-1 \textless 5  & Forest bird \\

27 & FC-2 \textgreater 4  & Heavy rain \\

28 & FC-2 \textless 5  & Forest bird  \\
 
29 & RST-out-IP \textgreater 5  and FC-5 \textless RST-out-IP and ACK-out-IP \textless 7 & Wind \\

30 & RST-in-IP \textgreater 5  and  FC-6 \textless RST-in-IP and ACK-in-IP \textless 7 & Wind \\

31 & SYN-ACK-out \textgreater 20   & Snow storm \\

32 & SYN-ACK-in \textgreater 20   & Walk in snow \\

33 & (Traffic Flow Counter) \textgreater 1000   & Fire \\

34 & (IP Flow Counter) \textgreater 600   & Fire \\
\bottomrule
\end{tabular}

\begin{flushleft} A selection of event conditions and their corresponding sounds.
\end{flushleft}
\label{table2}

\end{table*}

\subsection{SoNSTAR interactive sonification}

\Son\ is an interactive sonification system. Users may change the time window period, manipulate features and thresholds and re-assign sounds, and then restart with the new settings online. The level of each event sound can be adjusted independently with a slider control and can even be muted if desired. Any sound can be assigned to any chosen flow event in real time enabling the user to re-design the sound environment completely.

\begin{figure}[!h]
\includegraphics[scale=0.4]{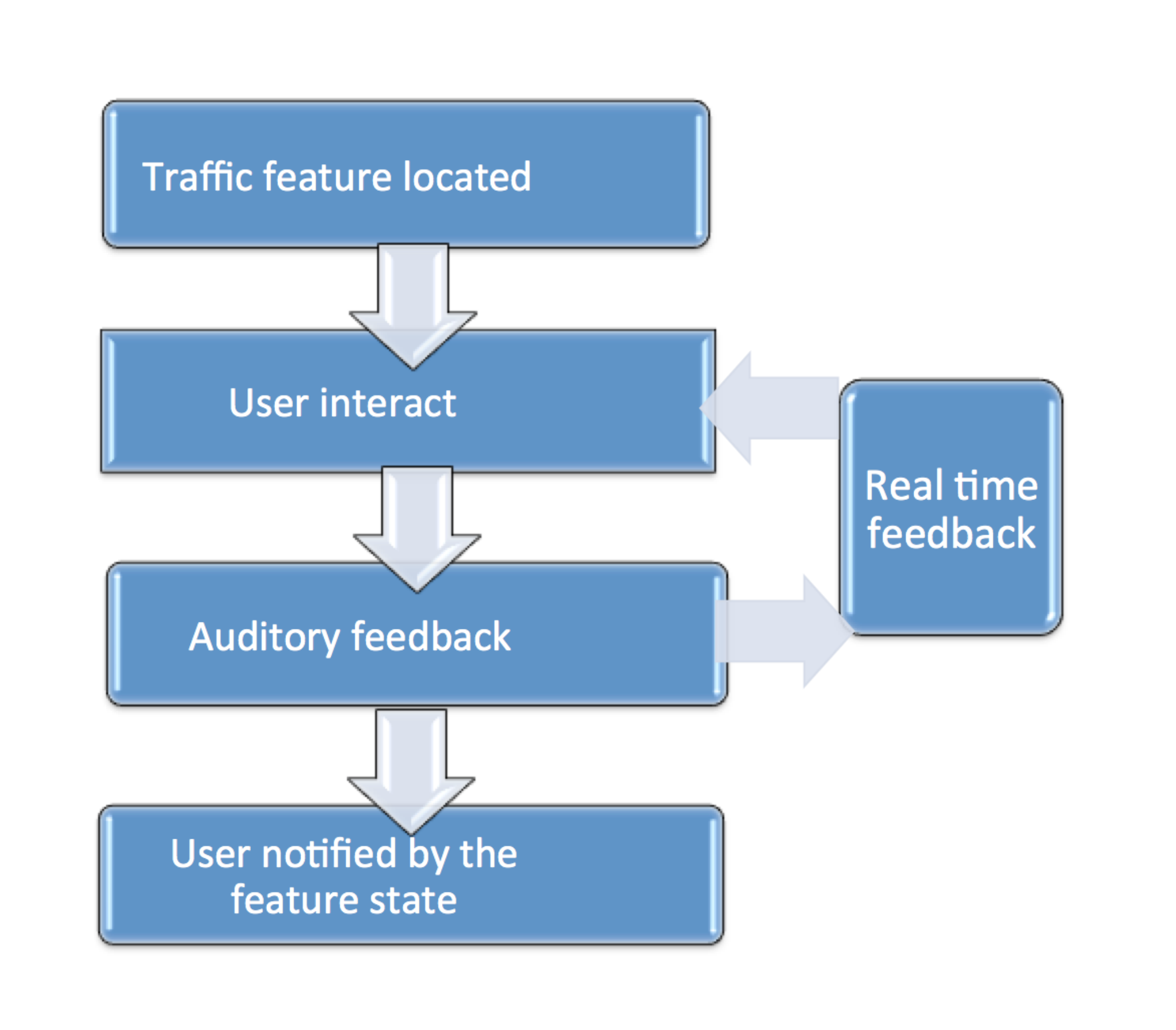}
\caption{{\bf Interactive sonification}
Model showing the interactive nature of the \Son\ sonification.}
\label{fig6}
\end{figure}

Users interact with \Son\ according to their understanding of the sound generated by the network traffic environment so as to increase their situational awareness. \Son\ enables the user to interact immediately with the system and its traffic to identify anomalous behaviours. Hunt and Hermann advise that sonification designers should respect `the bindings between physical actions and acoustic reactions that we have been familiar with since birth'~\cite[p. 295]{Hunt:2011}. In a network environment this could mean that we would expect the sounds to change when the system is under attack and we expect networks to behave differently when they are under more stress.  \Son\ uses multiple natural and man-made sounds to create the soundscape environment. When choosing the sounds, the natural reactions of users to the sounds is taken into consideration in order to allow users to sense and feel the network environment in relation to their own experience in the real world. \Son\ allows users to change sounds and create their own preferred acoustic environment in order to enable them to choose the most suitable sounds which  convey to them the state of the network in a maximally meaningful way.  \Son\ transforms all of the network traffic into a rich auditory field that envelops the listener in a goal-driven exploratory methodology where the network traffic is first filtered and the user is left only with the specific features that they chose.

\section{Experimental work and results}
A user study was conducted to investigate the monitoring of network behaviour by participants using \Son\ and, in particular, to evaluate \Son\ as a complement to existing system security tools. Three experimental conditions were investigated: 1) audio feedback only using \Son, 2) visual feedback only using the Snort intrusion detection software, and 3) audio and visual feedback together (\Son\ and Snort).  

\Son's  current design is able to extract TCP, UDP and ICMP protocol packet information. In this experiment only TCP and ICMP packet header information was extracted with ICMP packet data being used to detect ping activities. The sound of a woodpecker sound was assigned to ICMP ping activity.

\subsection{Network Design}

The experiment was conducted using two virtual networks running on the Virtualbox software. The first network was installed on a Mac OS 10.10.5 workstation with a 3.7 GHz quad-core processor, 16 GB 1866 MHz DDR3 ECC RAM and  a 27-inch (2560 x 1440) display. This virtual network comprised four machines (Ubuntu 64-bit, Windows Server 64-bit, Kali Linux Debian 64-bit and Mac OS 10.11) in addition to the host machine.

\begin{figure}[!h]
\includegraphics[width=\linewidth]{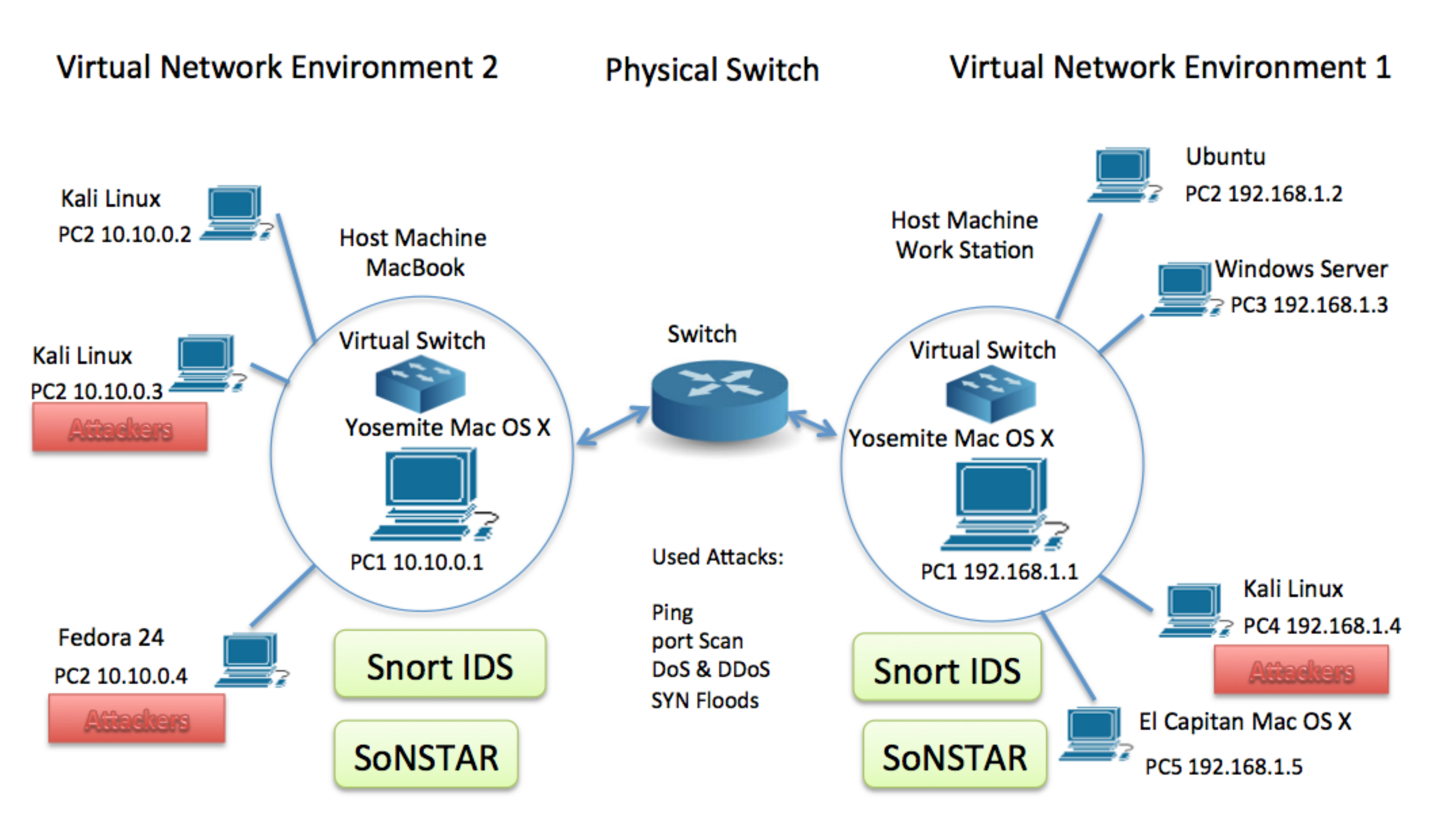}
\caption{{\bf Virtual network environment}
The virtual network environment design used in the experiment}
\label{fig7}
\end{figure}

The second virtual network was installed on a MacBook Pro running Mac OS  10.10.5 with a 2.5 GHz Intel core i7 processor, 16 GB 1600MHz DDR3 RAM and a 15.4-inch (2880 x 1800) Retina display. This network contained three machines (two Kali Linux Debian 64-bit installations and a Fedora 24 64-bit machine) in addition to the host machine.

These two virtual networks were connected through a router provided by Northumbria University. \Son\ and the Snort IDS were installed on both networks allowing each network to attack the other, and each machine to attack the other machines within its own local virtual network. 

\subsection{Participants}

A call for participants was been sent through the university email system to all MSc and PhD computer science and engineering students. 16 students responded to the email and 10 participants (7 male, 3 female) were able to devote the time needed to participate in the study which took place in September 2016. All 10 participants completed the study. All of the participants were aged from 25 to 45 years and 
 were PhD and MSc students at the university (8 from the Department of Computer and Information Sciences). All participants had good knowledge of the use of computers and information technology and general knowledge about computer network security. 
 
\subsection{Experimental design}

Each participant performed a network monitoring task under each of the three experimental conditions (audio only, visual only, audio-visual). Each task required participants to detect either 3 or 4 out of 7 attacks. 

The participants were assigned randomly to use Snort or \Son\ (five participants each) and then to use them together. At the end of each task performance was calculated based on the number of true positive (TP), true negatives (TN), false positives (FP)and false negatives (FN), where:

\begin{itemize}
 \item \textbf{TP:} the number of events which are correctly identified. The case was positive and was detected by the user as positive.
 \item \textbf{FP:} the number of events which are incorrectly identified. The case was negative but was detected by the user as positive.
 \item \textbf{TN:} the number of events which are correctly rejected. The case was negative and was detected by the user as negative.
 \item \textbf{FN:} the number of events which are incorrectly rejected. The case was positive but was detected by the user as negative.
 \end{itemize}

These variables are then used to calculate several metrics to assess the effectiveness of \Son\ as follows. 

The \emph{recall} metric (also known as the true positive rate) indicates the proportion of positives which are correctly detected by participants and is given by:

$$\text{recall}= \frac{\TP}{\TP + \FN}$$

\noindent The \emph{precision} is the number of true positives amongst all the reported positives:
$$\text{precision} =\frac{\TP}{\TP+\FP}$$

\noindent The F-measure is a weighted harmonic mean of the precision and recall~\cite [p.1147] {Liu:2009}:

 $$ F = 2\cdot\left(\frac{\text{precision}\cdot\text{recall}}{\text{precision}+\text{recall}}\right)$$

\noindent The \emph{accuracy} metric indicates  the proportion of correct identifications of all instances:

$$\text{accuracy}=\frac{\TP+\TN}{\TP+\TN+\FP+\FN}$$

\noindent The true negative rate (TNR) indicates the proportion of negatives that are correctly identified, such as the percentage of network events which are correctly identified as not occurred.

$$\TNR= \frac{\TN}{\TN + \FP}$$

\noindent The false positive rate (FPR) indicates the proportion of positives that are incorrectly identified, such as the percentage of network events which are incorrectly identified as occurred.

$$\FPR= \frac{\FP}{\FP + \TN}$$

\noindent The false negative rate (FNR) indicates the proportion of negatives that are incorrectly identified.

$$\FNR= \frac{\FN}{\FN + \TP}$$

 Snort's detection rules were set to the defaults provided by the \texttt{snort.conf} file. \Son\ was set to the sound mappings presented in Table \ref{table2}.

Four categories of behaviour were used in this experiment as follows: 

\begin{itemize}
 \item \textbf{Traffic:} using the internet, such as playing a YouTube video.  
\item \textbf{Ping:} using an ICMP ping.
\item \textbf{Port scan:} four types --- SYN, Null, Xmas and FIN port scans.  
\item \textbf{DoS, DDoS} including first, SYN flood as type; and second, DDoS using spoofed IP addresses performed from the three machines in the virtual network.
\end{itemize}

These behaviours were performed using a normal terminal, Nmap scanner and Hping3 commands. The supplemental material contains the files \texttt{S2-Audio}, \texttt{S3-Audio}, \texttt{S4-Audio}, \texttt{S5-Audio}, \texttt{S6-Audio}, \texttt{S7-Audio} and \texttt{S8-Audio} which are the \Son\ sonifications of the attacks used in this experiment.

\subsection{Materials}
\label{sec:mat}
Before beginning the experiment, each participant was given an informed consent declaration to sign (see \texttt{S5\_Appendix} in the supplemental material). Following the giving of consent each participant completed the three tasks using a Mac OS 10.10.5 workstation equipped with a 27-inch monitor and Sony  MDR-7506 Professional headphones. 

A questionnaire was given to each participant. The questionnaire can be found in the supplemental material (see \texttt{S2\_Appendix}). The first section elicited general  participant information such as sex, level of education, speciality, department and year of study.  The second section was a table for reporting detected malicious activities for the monitoring detection tasks for the three task conditions. The questionnaire provided two tick boxes in front of each type of attack for the three task conditions. 

The third section included evaluation of monitoring workload; upon completion of each experimental task participants completed the NASA-Task Load Index (TLX) assessment~\cite{Hart:1988} to measure their performance workload.  This includes mental demand, temporal demand, physical demand, performance, effort and frustration rates. Also there were extra ratings for detection confidence, ease of use, visual fatigue and sound fatigue included in the evaluation of both tools. For each of these rates, the participant had to provide an assessment rating on a scale of 0 to 10. 

The participants were then asked to choose their preferred condition (\Son, Snort, or both together). They were also requested to provide their evaluations of Snort and \Son\ on a scale of 0 to 5 where 5 denotes the most positive assessment. Participants could also provide feedback about this experiment in the final section.

The training and guidelines sheet included a table containing the seven chosen attack types for the experiment as well as the detection of text in snort and detection sounds in \Son\ written in front of each attack. The first column contained the attack category, the second column the attack type name, the third column text expected by Snort and the fourth column a description of the sound events for each attack, explaining the extra understanding those sounds provide. 

\subsection{Procedure}
Participants were informed that they would take the role of a network administrator to protect against malicious activities. The explanation of the experiment included three sections (one for each task condition) and where should they fill in the appropriate section for each task condition. The participants' virtual network computers were switched on and some music and YouTube videos were started to generate normal traffic across the network.

Participants were trained for about five minutes in the basics of the Snort IDS and another five minutes on \Son\ before starting each task condition. The rules for administration to protect their network and servers against attacks and malicious activities were explained including the seven specific attacks used in experiment. It was  also explained how each task condition involves concentration and high attention for long periods to detect attacks in their early stages. 

Training involved only the seven attack types used in this experiment. Participants were provided with a training and guidelines sheet and then trained on how Snort would show the detected attacks, and how Snort provides text warnings for each type. The seven attacks were demonstrated in real time. \Son\ training involved the same attacks but this time participants were provided with headphones and using the training and guidelines sheet they were asked to listen to the attacks one by one in real time. Any questions raised by participants were answered. They were not informed that \Son\ was a project under development so as to eliminate the effect of such knowledge on the results.

Each participant was provided with the questionnaire to fill in the outcomes for the three tasks. Five participants were assigned to the \Son\ condition for seven minutes first and then to the Snort  condition for another seven minutes. They were then assigned to use both \Son\ and Snort for another seven minutes.

The other five participants were assigned to the Snort condition for seven minutes  and then the \Son\ condition for another seven minutes. Then they were assigned to use both \Son\ and Snort for another seven minutes. This was done to eliminate the effect of using any one condition first.

During each period, the participants' networks received three or four real-time attacks. However, they were not informed about the number of malicious activities that could be expected. During each task, each participant was asked to continue speaking and were asked for more information about their understanding of security in order to affect their concentration to some extent. 

Directly after completing each task, participants had to answer the rest of the questions regarding the Monitoring Evaluation Tasks for each tool. At the end of the experiment, the participants were asked to tick which was considered the best for them to use, Snort or \Son\ or both together. Then they were requested to complete the rest of the questionnaire. 

\subsection{Results}

Several results are extracted from the questionnaire data as follows.

\subsubsection{TP, TN, FP and FN results}
    
The results for the three conditions are shown in Table~\ref{table3} as extracted from the questionnaire data. The results were calculated for the three conditions to assess \Son's capabilities  as part of the situational awareness process. Based on these results, various metrics are calculated to evaluate the \Son\ sound design and the usability of the system.

\begin{table}[!h]
\caption{TP, TN, FP and FN }
\centering
\label{tab:TF}
\begin{tabular}{llll}
\toprule
\textbf{Metrics} & \textbf{Snort} & \textbf{SoNSTAR} & \textbf{Snort and SoNSTAR} \\ 
\midrule
TP &  31 & 33 & 30  \\

TN &  31 & 33 & 38 \\

FP &   7 & 4 & 2  \\

FN &   0 & 0 & 0  \\
\bottomrule
\end{tabular}
\label{table3}
\end{table}

The metrics calculated from the base variables are shown in Table~\ref{table4}.

\begin{table}[!h]
\caption{Evaluation results}
\centering
\begin{tabular}{llll}
\toprule
\textbf{Metrics} & \textbf{Snort} & \textbf{\Son} & \textbf{Snort \& \Son} \\ 
\midrule
Recall &  100\% & 100\% & 100\%  \\

Precision &  81.58\% & 89.19\% & 93.75\%  \\

F-measure &  89.86\% & 94.29\% & 96.77\% \\

Accuracy &  89.86\% & 94.29\% & 97.14\%  \\

TNR &    81.58\% & 89.19\% & 95\% \\

FPR &   18.42\% & 10.81\% & 5\%  \\

FNR &   0\% & 0\% & 0\%  \\

\bottomrule
\end{tabular}

\label{table4}
\end{table}

The results show a maximum recall of 100\% for the three state conditions.  Meanwhile the TNR was higher when using \Son\ (89.19\%) compared to Snort (81.58\%). However, when participants used both together this rose to 95\%. The FPR was higher when using Snort (18.42\%) than \Son\ (10.81\%). However, when participants used both together this decreased to 5\%.

Accuracy was calculated for the three state conditions used in the experiment.  Accuracy of recognition was highest when using both Snort and \Son\ together at 97.14\%. \Son\ alone maintained higher accuracy than Snort alone, at 94.29\% and 89.86\% respectively. 

Precision was also calculated for the three state conditions used in this experiment.  Precision of recognition was highest when using both Snort and \Son\ together at 93.75\%. \Son\ alone maintained higher precision again compared to Snort at 89.19\% and 81.58\% respectively.

The F-measure was highest when using both Snort and \Son\ together at 96.77\%. \Son\ achieved a higher F-measure than Snort at 94.29\% and 89.86\% respectively. 

\subsubsection{NASA-Task Load Index results}

The NASA-Task Load Index results are shown in Table~\ref{table5}.

\begin{table}[!h]
\caption{NASA-Task Load Index results}
\centering
\begin{tabular}{llll}
\toprule
\textbf{No} & \textbf{Task Load Index} & \textbf{Snort} & \textbf{\Son}  \\
\midrule
1 & Mental Demand Rate & 58\% & 45\% \\

2 & Temporal Demand Rate & 65\% & 31\% \\

3 & Physical Demand Rate  & 28\% & 24\% \\
 
4 & Performance Rate &  82\% & 92\% \\

5 & Effort Rate  &  41\% & 19\% \\

6 & Frustration Rate  & 71\% & 36\%\\
\bottomrule
\end{tabular}

\label{table5}
\end{table}

\subsubsection{Additional evaluation results}

Additional \Son\ evaluation results are shown in Table~\ref{table6}.  

\begin{table}[!h]
\caption{Additional SoNSTAR evaluation (index results)}
\centering
\begin{tabular}{llll}
\toprule
\textbf{No} & \textbf{Task Load Index} & \textbf{Snort} & \textbf{\Son}  \\
\midrule
1 & Detection Confidence Rate & 88\% & 90\%\\

2 & Ease of Use Rate & 86\% & 96\% \\

3 & Visual or Sound Fatigue Rate  & 59\% & 40\% \\
\bottomrule
\end{tabular}
\label{table6}
\end{table}

Table \ref{table7} shows participants' opinions about whether using Snort and \Son\ alone or together would be best for monitoring.

\begin{table}[!h]
\caption{Additional SoNSTAR evaluation (preference results)}
\centering
\begin{tabular}{llll}
\toprule
\textbf{Index} & \textbf{Snort} & \textbf{SoNSTAR} & \textbf{Both together} \\
\midrule
Best to use & 10\% & 30\% & 60\%\\
\bottomrule
\end{tabular}
\label{table7}
\end{table}

Table \ref{table8} shows participants' opinions about Snort and \Son\ from horrible (H) to fantastic (F). 

\begin{table}[!h]
\caption{Horrible to Fantastic Evaluation }
\centering

\begin{tabular}{llllll}
\toprule
\textbf{Tool} & \textbf{H (100\%)} & \textbf{H (50\%)} & \textbf{Average} & \textbf{F (50\%)} & \textbf{F (100\%)} \\ 
\midrule
Snort &  0 & 0 & 40\%  & 10\% & 50\%\\

\Son &   0 & 10\% & 30\%  & 0\% & 60\% \\

\bottomrule
\end{tabular}
\label{table8}
\end{table}
 
The most remarkable feedback was that a participant advised that if it is possible to add a visual panel showing the name, colour and image  helps to distinguish the recorded sound in order to facilitate learning and confirmation, which facilitates cognitive process and get used to link events with the recorded sound.

\section{Discussion}
These experimental results clearly show improvements in monitoring when using sonification compared to the visual method only.  Although the detection rate was 100\% for the three state conditions, we can still see  improvements in the accuracy, precision and F-measure scores for the sonification conditions.  Although the training of participants was very brief and the computer security background of most of the participants was basic, they were able to use both systems well in a short time.

\Son\ can fill the gap between the network operator and the traffic environment by providing an auditory link between traffic behaviour and the operator's mind. The experiment showed that users with a basic knowledge of computer network principles can use \Son\ to learn how  flows are behaving inside network traffic and to recognise the typical combinations of the packet types within the flows. To ease the learning process, \Son\ is generates log files holding counts of the packets within flows in addition to messages that indicate which flow event generated which specific sound. These reports support the user in understanding the links between flag states, packet counts, flow counts, and the resultant sequences of sounds. Therefore, after learning the basics of network the TCP/IP protocol, \Son\ users can examine network behaviours; indeed, \Son\ could be used as an instructional tool to help educate new users computer networks. The IP-flow  and traffic-flow log files together can explain in detail how the packet counts of the two flow types are related (see  \texttt{S3\_Appendix} IP flow log file and  \texttt{S4\_Appendix} Traffic flow log file in the supplemental material).  The project repository \cite{nuson-SoNSTAR} contains the source code and Max/MSP patches necessary to install and run SoNSTAR together with sample data files, example output stored in audio files, and instructions on how to use the system.

Some behaviours will create multiple sounds. For example, when performing a SYN scan, the attacker will send a number of packets with the SYN flag set to 1 to a number of targeted ports. If the port is open the receiver would send back a packet with the SYN flag set to 1 and the ACK flag set to 1 as a reply to accept the connection. The attacker either sends back a packet with the FIN flag set to 1 to cut the connection (the TCP half-open scan type) or sends two packets, the first with the ACK flag set to 1 to confirm the connection and then the second packet with the FIN flag set to 1 to cut the connection (the TCP connect scan type). If the port is closed the receiver would send back a packet with the RST flag set to 1, and if there is no response it means that the port is filtered.

As we set \Son\ to default settings, as soon as it receives many packets of the SYN type in an IP flow, \Son\ will play the rain-on-a-roof sound and this would tell the user that an unusual number of SYN packets is arriving. If the TCP handshake was not correct, that event would generate a heavy rain sound which would tell the user that there is a problem with connecting to a specific IP address. If the number of SYN packets was high, \Son\ will play a thunder sound and this will tell the user that someone is scanning a large number of the system ports of a specific IP address. If the number was huge it would be considered a DoS attack and the sound of fire would be played. If the scanned system started to send out RST packets, \Son\ would play the sound of wind, confirming that it is a scan attack. This is a complex process, but \Son\ would deal with any changes in behaviour and play sets of sounds according to what events are happening in the network. The user could identify any new behaviour according to the set of sounds played. 

Using such a tool to explore and tune a network is important due to the different nature of networks and the different expected behaviours with different thresholds. For example, this tool could be used to tune IDS settings to look for new features and events which could be used to identify threats on a particular network. Using \Son\ to draw a normal base line for a specific network behaviour would help to make rules and thresholds for specific environment which will raise the situation awareness in general.    

Another advantage that \Son\ provides is that it generates log files which could help any user to learn and confirm how are packet types counts of IP flow effects recorded sound played and to evaluate theoretically any event and ideas of any new feature (Review  \texttt{S3\_Appendix} IP flow log file which contains SYN scan in the supplemental material). \Son\ could help users and network students to explore network protocols and to learn more about network traffic. The use of SoNSTAR would enable them to think directly about the logic of any behaviour in network traffic and would give them the opportunity to express their own ideas and to test and learn from them. Using \Son\ reduced mental demand, temporal demand, effort and frustration rates significantly compared to using Snort (visual tool) and this would be more obvious if the monitoring was for long hours. This confirms that the use of \Son\ increases the situational awareness of the operator as it not only gives a warning of the attacks, but gives more than that so that the operator listens to what happens to the network directly so that the operator links between behaviours and sounds, this relationship enables him to understand what is happening accurately by distinguishing the sounds and its meaning to allow comprehension which enables him to make decisions directly without reference to any alarms reporting or going through long forensic process to understand the behaviour. This seen clearly from the increase of detection confidence rate when using \Son\ and this could more noticed if the monitoring period was longer.   

\section{Conclusion and further work}

This study indicated that using sonification improved the monitoring process, even for people who have only basic knowledge about network monitoring. Using sound reduced the overall mental work load. Participants were able to recognise and comprehend behaviours and decide which attack was performed which proved human mind could learn quickly about the network environment in a way would result in increasing the security situational awareness. Although the system could be evaluated manually by comparing against the log files, this experiment evaluated the practicality of using sonification in live monitoring tasks. The results suggest that using \Son\ to explore new event and features would bring benefits to IDS systems and network monitoring in general and for situational awareness.

From \Son\ research one might get the impression that there is much more work that can be done to evaluate and improve the attack detection and improve the situational awareness. we have been looking at feature extractor and combiner and we succeeded to develop many features while finding them can be a difficult task, this features could be submitted to the publicly open databases, were people could obtain these feature to develop their detection systems. Developing a language for describing these discovered features and events based on network traffic with presenting the log files explaining how these feature can be used could be some of the future work in this area.

The major contribution made to reduce the complexity of huge volumes of traffic in order to be comprehensively sonified  by using IP flow in detecting network behaviour, especially to horizontal behaviours, Also the sound mapping of the network event based on packet type counts have not been seen before.  We suggest further work to create more features in order to target vertical behaviours.
     Other important future work consists of developing a method to create features for UDP, ICMP, IRC and other protocols using \Son\, so that larger amounts of representative traffic is covered for developing and testing of network traffic. Also \Son\ has very high potential to represent SCADA system because of the unique nature of them. If their normal behaviour footprint are tested a zero day attack can be guaranteed. Further study could be performed to evaluate what could \Son\ provide to the educational process to enable the students to see traffic in simple and meaningful way as contribution. 

\section{Supporting information}
\subsection{Documents}
The following documents are available from \url{https://github.com/nuson/SoNSTAR/tree/master/docs}.
\begin{table}[!h]
\caption{Supporting Documentation}
\centering
\label{tab:doc}
\begin{tabular}{L{3cm}L{5cm}}
\toprule
\textbf{Filename} & \textbf{Description} \\ 
\midrule
\texttt{S1\_Appendix.pdf} & SoNSTAR: flow and IP flow feature information array contents. This file shows the contents of the feature information arrays for Traffic flows and IP flows\\
\texttt{S2\_Appendix.pdf} & Experiment questionnaire. This file shows the contents of the questionnaire used in this experiment for evaluation.\\
\texttt{S3\_Appendix.pdf} & SoNSTAR IP flow log file. Text file contains IP flows information. Each row contains time window number, then the IP flow number within current time window, then host A IP address, host B IP address, and then the feature counts sorted in the same sequence in \texttt{S1\_Appendix}\\
\texttt{S4\_Appendix.pdf} & SoNSTAR Traffic flow log file. Text file contains traffic flow information. Each row comprises time window number, then the Traffic flow number within current time window, then host A IP address, host B IP address, , then host A port number, host B port number, and then the feature counts sorted in the same sequence in\texttt{S1\_Appendix}\\
\texttt{S5\_Appendix.pdf} & The informed consent form. This file shows the consent form used in this experiment.\\
\bottomrule
\end{tabular}
\end{table}

\subsection{Audio}
The following audio files are available from \url{https://github.com/nuson/SoNSTAR/tree/master/examples}.
\begin{table}[!h]
\caption{Supporting Documentation}
\centering
\label{tab:doc}
\begin{tabular}{L{3cm}L{5cm}}
\toprule
\textbf{Filename} & \textbf{Description} \\ 
\midrule
\texttt{S1-Audio.aif} &  Normal traffic behaviour.  \Son\ normal events sounds audio file.\\
\texttt{S2-Audio.aif} &  FIN behaviour. \Son\ FIN scan audio file. Scan performed using hping3.\\
\texttt{S3-Audio.aif} &  Xmas behaviour. \Son\ heavy Xmas scan audio file. Scan performed using Nmap. \\  
\texttt{S4-Audio.aif} &  NULL behaviour. \Son\ low NULL scan audio file. Scan performed using hping3.\\
\texttt{S5-Audio.aif} &  NULL behaviour. \Son\ heavy NULL scan audio file. Scan performed using hping3.\\
\texttt{S6-Audio.aif} &   SYN behaviour.  \Son\ heavy full connection SYN scan audio file. Scan performed using Nmap.\\
\texttt{S7-Audio.aif} &  Ping behaviour.  \Son\ SYN-Flood-DOS audio file. sounds of SYN flood attack behaviour for denial of service purpose. Performed using hping3.\\
\texttt{S8-Audio.aif} & Ping behaviour.  \Son\ Null-DDOS audio file. DDOS (distributed denial of service) using null packet type. Performed using hping3.\\
\bottomrule
\end{tabular}
\end{table}


\end{document}